\documentstyle[11pt,paspconf,psfig]{article}
\begin{document}

\title{Infrared observations of cataclysmic variables}
\author{Vik Dhillon}
\affil{Royal Greenwich Observatory, Madingley Rd, Cambridge CB3 0EZ, UK}

\begin{abstract}
We review   infrared  (1--2.5  $\mu$m)  observations   of  cataclysmic
variables, a relatively  unexplored part of the  spectrum in which the
dominant sources of emission are the secondary star, the outer regions
of the accretion disc and the accretion column in magnetic systems. We
describe the   advances that have  been  made in our  understanding of
cataclysmic variables based on infrared photometry and, more recently,
infrared spectroscopy and present spectra of each class of cataclysmic
variable  -- the dwarf  novae,    novalikes, polars and   intermediate
polars. 
\end{abstract}
 
\section{Introduction}

The infrared (IR) extends from a wavelength of 1  $\mu$m, close to the
limit of  optical CCDs, through the   near-infrared (1--5 $\mu$m), the
mid-infrared (5--25 $\mu$m) and the far-infrared (25--350 $\mu$m).  In
contrast to  the optical,  the molecules of  H$_2$O and  CO$_2$ in the
earth's  atmosphere absorb  very   strongly at  some IR   wavelengths,
resulting in relatively  transparent   regions  which define  the   IR
photometric      bands:     J~(1.1--1.4~$\mu$m),  H~(1.5--1.8~$\mu$m),
K~(2.0--2.4~$\mu$m),    L~(3.0--4.2~$\mu$m),      M~(4.5--5.2~$\mu$m),
N~(8--13~$\mu$m) and  Q~(17--22~$\mu$m). At wavelengths longer than Q,
the  IR is  only  accessible  from  space.  The  vast majority  of  IR
observations of cataclysmic variables (CVs) have been performed in the
so-called non-thermal part of  the near-infrared (1--2.5 $\mu$m),  and
for good reason.  This is the only IR region where  CVs can be readily
detected  against the background,    as thermal emission  due  to  the
observing environment (such as the telescope structure and optics) and
from  optically thick telluric absorption lines  rises very rapidly at
wavelengths  longward of  2.5 $\mu$m, to  such an  extent   that at 10
$\mu$m the background  is   approximately equal in  brightness  to the
brightest astronomical  source  known.  The 1--2.5  $\mu$m  wavelength
range  also  happens   to be  where  the  spectrum  of the  G--M-dwarf
secondary  star in CVs   is expected to  peak\footnote{The wavelength,
$\lambda_{max}$, of the peak flux, $f_{\nu}$, for  a star of effective
temperature        $T_{e\!f\!f}$    can    be     approximated     by:
$\lambda_{max}=5100/T_{e\!f\!f}$~$\mu$m.  With   $T_{e\!f\!f}$ ranging
from  $\sim$6000--2000~K on the  G--M-dwarf   secondary stars in  CVs,
$\lambda_{max}$      ranges   from   $\sim$1--2.5~$\mu$m.}   and where
low-harmonic,  cyclotron emission from  the weaker-field  magnetic CVs
and emission from  the cool, outer  regions of  the  accretion disc in
non-magnetic CVs would be expected to  fall. For the above reasons, it
is  clear  that observations of  CVs in  the 1--2.5~$\mu$m  region are
highly desirable and this review  will concentrate exclusively on this
wavelength range -- for a description of  the very few observations of
CVs  at longer IR wavelengths, which  are mainly sensitive to emission
from dust grains in CVs, see Berriman, Szkody \& Capps (1985), Jameson
et al. (1987),  Harrison \& Gehrz  (1992) and Howell, Herzog \& Robson
(1996). 

Observing in the 1--2.5 $\mu$m region is not easy.  There are two main
reasons for this. First, even though this is the non-thermal region of
the IR, the background is still  high compared to  the optical part of
the spectrum.   This is  due  to  atmospheric emission  from molecules
(primarily OH$^-$ and O$_2$)  excited  by solar radiation during   the
day. The  resulting airglow is  both temporally and spatially variable
and results in a sky which can be up to 1000  times brighter in K than
it is in V. Second, IR detector arrays  have only recently come of age
and still have higher readout noise, higher  dark current and are of a
smaller size than their optical counterparts.  It is for these reasons
that   the  astronomical literature on   IR   observations  of CVs  is
relatively sparse  when  compared with optical, ultraviolet  and X-ray
studies  (with the  exception of  novae, which  are reviewed by  Gehrz
elsewhere in this volume and will not be  discussed further). In fact,
one might regard the present status of IR observations  of CVs as akin
to the status of optical observations of CVs nearly 40 years ago, when
Kraft  and co-workers were performing  the first spectroscopic surveys
and time-resolved studies of CVs in the optical. 

\section{Infrared photometry of cataclysmic variables}

The first IR observations of CVs were performed  by Szkody (1977), who
obtained J,  H,  K  and L-band photometry   of  a number  of CVs,  and
Sherrington et al.  (1980), who obtained  the first IR light curves of
CVs (EX Hya and VW  Hyi in the J  and K-bands). There followed a burst
of observational activity, culminating in the work of Berriman, Szkody
\&  Capps  (1985), who   reviewed the  origin  of the  IR   light from
CVs.  They  found that the  dominant   contributors are the (optically
thick) atmosphere  of the secondary  star and both the  optically thin
and the (cool)  optically thick outer  regions of  the accretion disc,
the optically thin emission arising from the same  gas that gives rise
to   the  optical and   ultraviolet   emission lines. By   plotting IR
two-colour diagrams, Berriman, Szkody and  Capps (1985) found that the
proportion of  light supplied  by  each  component varies widely  from
system to system due to the complex nature of the disc.  This makes it
impossible   to disentangle the   secondary   star and accretion  disc
components by IR photometry alone, as an earlier-type secondary and an
optically thin disc can  account for the IR colours   of a CV just  as
well  as a later-type secondary and  an opaque disc. As a consequence,
IR photometry can only give an upper limit  to the proportion of light
contributed by the secondary star. 

IR photometry of  CVs has proved of  most use in  the determination of
two  fundamental parameters: the inclination of  the orbital plane and
the distances to CVs. In a binary of moderate to high inclination, the
asymmetric Roche  lobe   causes the  flux  from  the secondary  to  be
modulated twice  per orbital period,  with maximum brightness when the
lobe is  seen sideways on.  Apart from  directly confirming the highly
distorted  shape   of  the   secondary,  this   so-called  ellipsoidal
modulation  can aid in   determining  the  inclination of  the  binary
(e.g.   Berriman et  al.  1983;  Somers,  Mukai \&  Naylor 1996).  Any
contribution to the IR light from the accretion disc, however, results
in an observed modulation which is smaller than the actual ellipsoidal
modulation.   As the    amplitude  of the  ellipsoidal  modulation  is
correlated with the orbital inclination (large amplitudes imply a high
inclination), this  means that modelling the  observed IR light curves
will underestimate  the binary inclination.  The only way of obtaining
accurate inclinations in this way is  to determine the contribution of
the accretion disc to the IR light by  detecting absorption lines from
the secondary star with IR spectroscopy (e.g. Shahbaz et al. 1996). 

The  distances to CVs can be  measured from  K-band photometry using a
method first  proposed by Bailey  (1981) and later refined by Berriman
(1987) and  Ramseyer (1994). The distance  modulus can be rewritten in
terms of the K-band  surface  brightness of  a star  as $S_k=K+5-5\log
d+5\log (R/R_{\odot})$, where $K$ is  the apparent K magnitude, $d$ is
the distance in parsecs and $R$ is the radius  of the star. For a star
of  one  solar  radius,   $S_k$ is   equivalent   to the   absolute  K
magnitude.  Since the radius of  the  secondary star  is equal to  the
radius of the Roche lobe, the orbital period and mass of the secondary
are sufficient to estimate its radius (there is also a weak dependence
on mass ratio). Given the  K magnitude, all of which  is assumed to be
due to the secondary, and the value of $S_k$, which can be obtained if
one knows  the V--K colour (or spectral  type)  of the secondary using
the  empirical calibrations  derived   from field dwarfs  by  Ramseyer
(1994), it   is  possible to estimate the    distance using  the above
equation. This technique  has  been put to good   use by a   number of
authors, such as Warner  (1987), who used  the distances to  determine
the  absolute magnitudes of the discs  of CVs,  and Sproats, Howell \&
Mason (1996), who used the distances  to determine that most faint CVs
at high galactic latitude are not in  the galactic halo but are nearby
and intrinsically  faint.  It  was   argued above,  however, that   IR
photometry can  only  ever give an  upper  limit to the proportion  of
light   contributed  by the  secondary star   and  hence the distances
determined by this method are only ever  lower limits. The only way of
obtaining accurate distances is  to determine the contribution  of the
accretion disc to the K-band  light by detecting absorption lines from
the secondary star with IR spectroscopy. 

\section{Infrared spectroscopy of cataclysmic variables}

\newcommand{\gta}{{\small\raisebox{-0.6ex}
{$\,\stackrel{\raisebox{-.2ex}{$\textstyle >$}}{\sim}\,$}}}
\newcommand{\lta}{{\small\raisebox{-0.6ex}
{$\,\stackrel{\raisebox{-.2ex}{$\textstyle <$}}{\sim}\,$}}}

It has only recently become possible to perform IR spectroscopy of CVs
at              useful               spectral              resolutions
($\lambda$/$\Delta\lambda\gta$10$^2$--10$^3$):  the first published IR
spectra of CVs were of the intermediate polar EX Hya (Bailey 1985; see
also the very low  resolution spectrum of Frank et  al. 1981) and  the
polar   AM Her (Bailey, Ferrario  \&  Wickramasinghe 1991).  The first
spectral surveys in  the IR were published by  Ramseyer et al. (1993),
at a low spectral resolution, and Dhillon \& Marsh (1993, 1995), at an
intermediate spectral resolution.  To    date,  there have been     no
published time-resolved studies or high spectral-resolution studies of
CVs in the  IR.  To further  illustrate how rarely these  objects have
been studied in the IR it should be noted that,  with the exception of
the work mentioned above and the work on polars by Ferrario, Bailey \&
Wickramasinghe (1993,  1996), every IR spectrum  of a CV that has ever
been published (to the best of the author's knowledge) is presented in
Figures~1--5 and is described below, grouped by class of CV. 

\subsection{Dwarf novae above the period gap}

\begin{figure}
\psfig{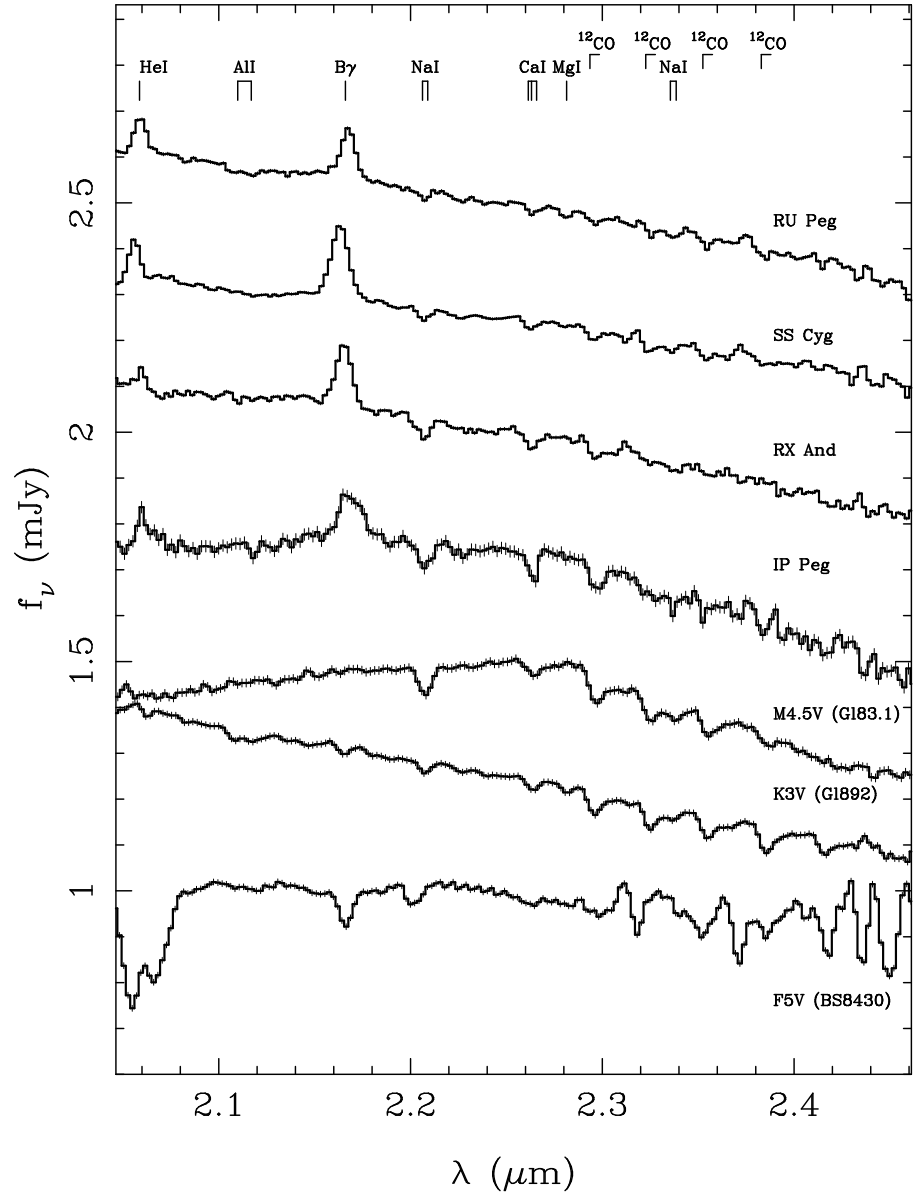}
%\vspace{6.6in}
\caption{Infrared spectra  of the dwarf novae  RU Peg, SS Cyg,  RX And
and  IP Peg, which all lie  above the period   gap. Also shown are the
spectra of  M4.5 and K3 dwarf  stars. The spectra have been normalized
by dividing by  the flux at  2.24~$\mu$m and  then offset  by adding a
multiple of 0.25 to each spectrum. The lowermost spectrum is of an F5V
star, normalized by  dividing by a  spline fit to its continuum, which
indicates the location of telluric absorption features.} 
\end{figure}

In Figure~1 we  present IR spectra of  dwarf novae which lie above the
$\sim$2--3~hr period gap (RU  Peg, SS Cyg, RX And  and IP Peg; Dhillon
\&  Marsh 1995). The IR  spectra of the  dwarf novae  are dominated by
strong emission lines  of He\,{\small I}  and the Paschen and Brackett
series of  H\,{\small I}.  The large  velocity  widths of these  lines
(typically 1000--2000 km\,s$^{-1}$ FWHM,  depending on the inclination
of  the binary) and  the fact that   they are so  strongly in emission
(typical equivalent widths of  $\sim$10\,\AA) imply an accretion  disc
origin.   The  IR spectra of the  dwarf  novae also   show a number of
absorption features  which, on comparison with the  IR  spectra of the
M4.5V and K3V field-stars in  Figure~1, can be identified with neutral
metal lines  of   Al\,{\small  I},  Na\,{\small  I},   Ca\,{\small I},
Mg\,{\small   I} and  $^{12}$CO  molecular-bands   from  the secondary
star. The    M4.5V  field-star  also exhibits    a   distinctive water
absorption band longward of $\sim$2.3~$\mu$m, which  is not present in
the K3V field-star. This feature is an excellent indicator of spectral
type in CVs as  it is so  prominent. As one  might expect, the  longer
period CVs in Figure 1 (RU Peg, SS Cyg, RX And) do not show this water
band,   implying their secondary  is  a  K-star, whereas the  shortest
period CV in Figure 1 (IP Peg) does show  the water band, implying its
secondary is an M-star. 

Once absorption features have  been detected, there   are two ways  in
which  the contribution of  the secondary star to the  IR light can be
determined  from IR spectra. The   first is by an optimal  subtraction
technique (Dhillon \& Marsh  1993), where a  constant times a spectrum
of  a field-dwarf  template is  subtracted from  the CV spectrum,  the
constant  being chosen to  minimize  the residual  scatter  in regions
containing  secondary star  features    (the scatter is   measured  by
carrying out the subtraction  and then computing the $\chi^2$  between
the resulting spectrum and a smoothed version of itself). The spectral
type of the template  which gives the lowest value  of $\chi^2$ is the
spectral type of  the secondary star.  Note that if  the above optimal
subtraction  is performed on  normalized spectra, the constant is then
equal to  the fractional contribution of  the secondary star to the IR
light  and the    resulting  spectrum  is   that of   the   fractional
contribution of  the accretion disc.  The  second  technique relies on
plotting  flux-deficit  ratios (Wade \&  Horne 1988;  Dhillon \& Marsh
1995).  The equivalent widths of individual absorption features cannot
be used to determine the contribution of the  secondary star to the IR
light,  since they  are affected  by the  continuum from the accretion
disc. The accretion disc is too hot to contribute  to the cool stellar
absorption features, however,  and hence the ratios of absorption-line
flux deficits   can be used.  In the  K-band,  Dhillon \& Marsh (1995)
found  the   optimum results were  obtained  using   the  ratio of the
strength of the water band longward of $\sim$2.3 $\mu$m to Na\,{\small
I} at 2.21 $\mu$m plotted against the ratio of the  same water band to
the  predicted continuum level above the  water band. On  such a plot,
the spectral-type of a CV can be determined from its position relative
to field-dwarf templates on    the water/Na\,{\small I} axis and   the
contribution of the disc to  the IR light  is given by the distance of
the  CV  from  the field-dwarf   templates  along the  water/continuum
axis. From the spectra presented  in Figure~1, Dhillon \& Marsh (1995)
concluded that the spectral   types of the secondary stars  determined
from IR   spectra are  in agreement    with  optical estimates,  which
indicate  that all but  the   longest  period CVs  generally   contain
secondary stars which are  indistinguishable from  main-sequence stars
(Smith \& Dhillon 1997). 

\subsection{Dwarf novae below the period gap}

\begin{figure}
\psfig{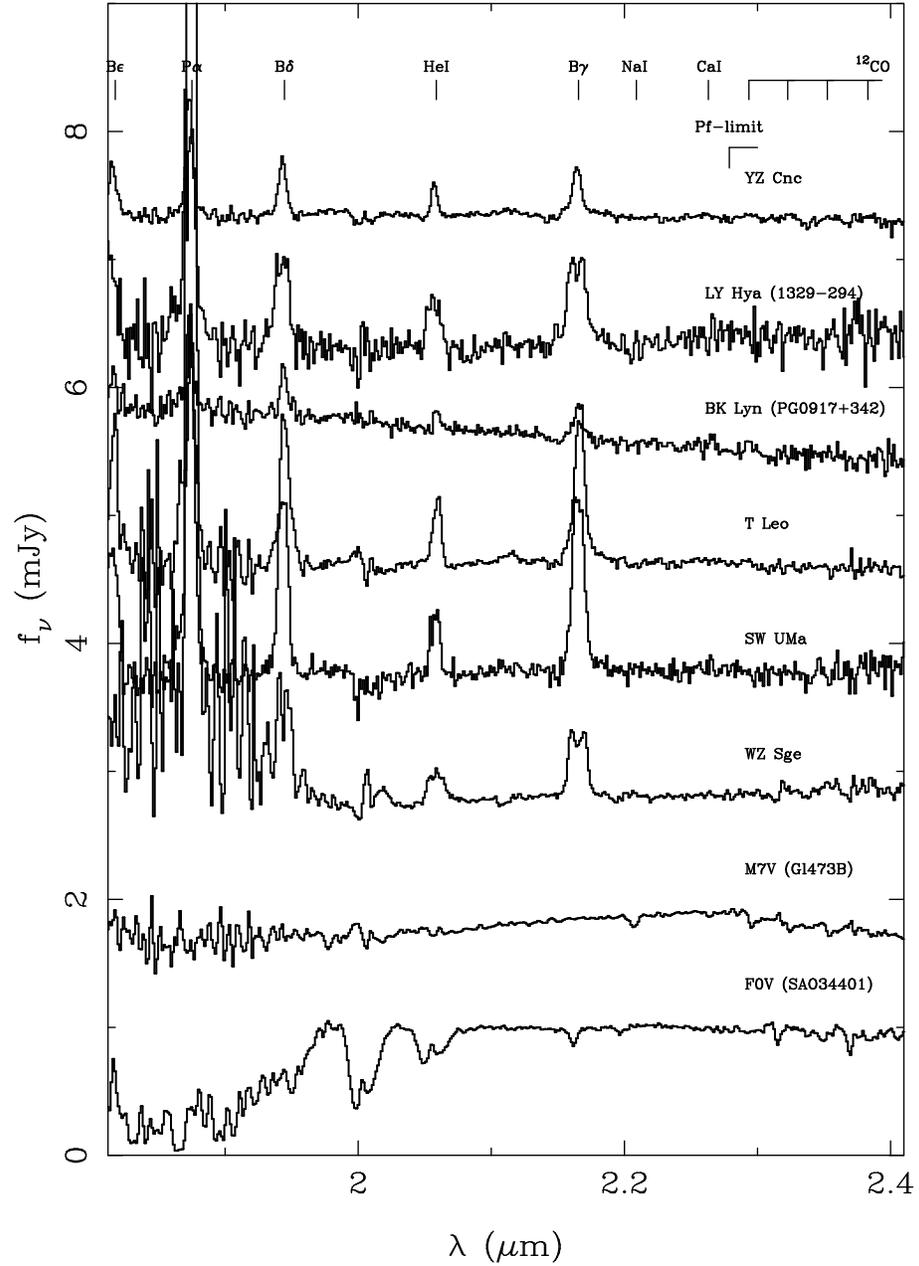}
%\vspace{6.6in}
\caption{Infrared spectra of the dwarf novae YZ Cnc, LY Hya, BK Lyn, T
Leo,  SW UMa and WZ Sge,  which  all lie below   the period gap.  Also
shown is  the spectrum  of an M7   dwarf star.  The spectra have  been
normalized by  dividing by the flux at  2.24~$\mu$m and then offset by
adding a multiple  of 0.9 to each  spectrum. The lowermost spectrum is
of  an F0V  star,  normalized by   dividing  by  a spline fit   to its
continuum, which  indicates    the  location of  telluric   absorption
features.} 
\end{figure}

In Figure~2 we present IR  spectra of dwarf novae  which lie below the
period gap (YZ Cnc, LY Hya, BK Lyn, T Leo, SW  UMa and WZ Sge; Dhillon
et al.  1997b). As in  Figure 1, the spectra  are dominated  by strong
emission   lines  of H\,{\small  I}  and   He\,{\small I}.  The higher
resolution  of this data (compared to  the data presented in Figure 1)
has resolved the double-peaked emission  lines of the high-inclination
dwarf novae  LY Hya and WZ Sge,  thereby confirming that  the emission
lines  originate in the accretion disc.  Unlike the  IR spectra of the
dwarf  novae   above the  period  gap,  however,   there are no  clear
detections of absorption  features from the secondary  star in the  IR
spectra of the dwarf  novae below the period   gap. The only  possible
exception is WZ    Sge, which appears  to  display  the red  continuum
indicative of a late-type secondary star (c.f. the spectrum of the M7V
star at   the foot of  Figure~2).  If  the continuum   in WZ  Sge were
largely due to  the secondary star, however, one  would also expect to
see a change  in slope redward of $\sim$2.3~$\mu$m  due to  the strong
water absorption band   displayed   by M-dwarfs.   Such   a  change in
continuum slope is not observed in WZ Sge,  although it is conceivable
that blended emission lines near the H\,{\small  I} Pfund-limit act to
fill in the absorption band. 

Most of  the dwarf novae presented  in Figure~2 have also been studied
in the optical for evidence of  absorption features from the secondary
star --  in every case   the secondary remained  undetected (Friend et
al. 1988; Still et al. 1994; Smith et al. 1997).  In fact, the results
of these optical  surveys reveal that virtually  all  of the secondary
stars in dwarf novae  below the period gap  remain undetected (HT  Cas
(Marsh 1990)  and Z  Cha (Wade  \& Horne  1988) are  two of  the  rare
exceptions to this rule). The secondary  stars in CVs below the period
gap    are  expected    to    be  of   spectral    type   $\sim$M5  or
later,\footnote{The spectral type of the secondary star in a CV can be
determined  from its  orbital  period,  $P$, using the   relationships
$27.1-0.9 P$  (for  $P<4$~hr) and $33.7-2.7  P$  (for $P>4$~hr), where
G0=0, K0=10 and M0=20 (Smith \& Dhillon 1997).}  indicating that their
spectra should peak in the K-band. Hence  it is perhaps not surprising
that optical surveys have struggled to  find the small, cool secondary
stars in dwarf  novae  below the period  gap.  The fact  that even our
K-band survey presented in Figure~2 has not detected them implies that
the secondary stars in dwarf novae  below the gap generally contribute
a smaller fraction (typically $<$25\%) of the  total IR light than the
secondary stars  in dwarf novae   above the period gap  (which usually
contribute $>$75\%). 

There  is a strong  motivation to  detect the  secondary stars in  CVs
below the period gap. The models  of Kolb (1993)  predict that 99\% of
the present-day intrinsic population of CVs should be below the period
gap and approximately 70\% of these systems  will have already reached
the orbital  period minimum (at $\sim$80 min)  and  should be evolving
back towards  longer   periods.     The secondary stars    in    these
post-period-minimum CVs  have  been modelled  by  Howell, Rappaport \&
Politano (1997),  who find them   to  be degenerate, brown  dwarf-like
objects  with masses between  0.02--0.06~$M_{\odot}$  and  radii  near
0.1~$R_{\odot}$.    Howell,   Rappaport \&   Politano  (1997)  further
speculate that  the   so-called  tremendous outburst   amplitude dwarf
novae, or TOADs, are  these post-period-minimum CVs.  If this is true,
then one  test of their hypothesis would  be  to obtain IR  spectra of
TOADs with orbital periods near 2~hr (the longest period attainable by
a  post-period-minimum CV given current  estimates  for the age of the
galaxy) and determine the spectral types  of their secondary stars. If
the TOADs are pre-period-minimum CVs, a  system with an orbital period
of 2~hr  should contain  a main-sequence   secondary of spectral  type
$\sim$M5,$^2$ whereas if the TOADs are post-period-minimum CVs, a 2~hr
system  would have a  brown   dwarf-like  secondary, with a   spectrum
similar to that of a very late-type M-dwarf  (cooler than M9; Jones et
al. 1994).  Assuming  the  secondary star  can be  detected,  it is  a
simple matter to differentiate between  an M5 and an  M9 dwarf with IR
spectra, thanks  to the strong water absorption  bands  around 1.7 and
2.3~$\mu$m, which   show  a  dramatic  increase in   strength  towards
later-type M-dwarfs  (Jones   et al.  1994).   The  problem  with this
approach, of course, is  that it is not easy  to detect the  secondary
stars in CVs below  the period gap, as  evidenced by the IR spectra of
the TOADs WZ Sge, T Leo and SW UMa displayed in  Figure~2.  Even if we
had detected the secondary star in WZ  Sge, T Leo  or SW UMa, however,
these TOADs all have  periods close to  the orbital-period minimum and
are hence not very sensitive to the test outlined above. 

If one were to find post-period-minimum CVs, this  would have a number
of very interesting  consequences. First, it  would allow one to study
brown  dwarf-like  secondary stars,  and   by invoking  the Roche-lobe
filling criterion,  investigate the mass-radius relation.  Second, the
orbital periods of CVs with brown dwarf secondary stars can be used in
conjunction with the models of Howell, Rappaport \& Politano (1997) to
place a lower limit  on the present-day age  of the Galaxy, $t_G$, and
therefore an upper limit to the Hubble constant, $H_0$. 

\subsection{Novalikes}

\begin{figure}
\psfig{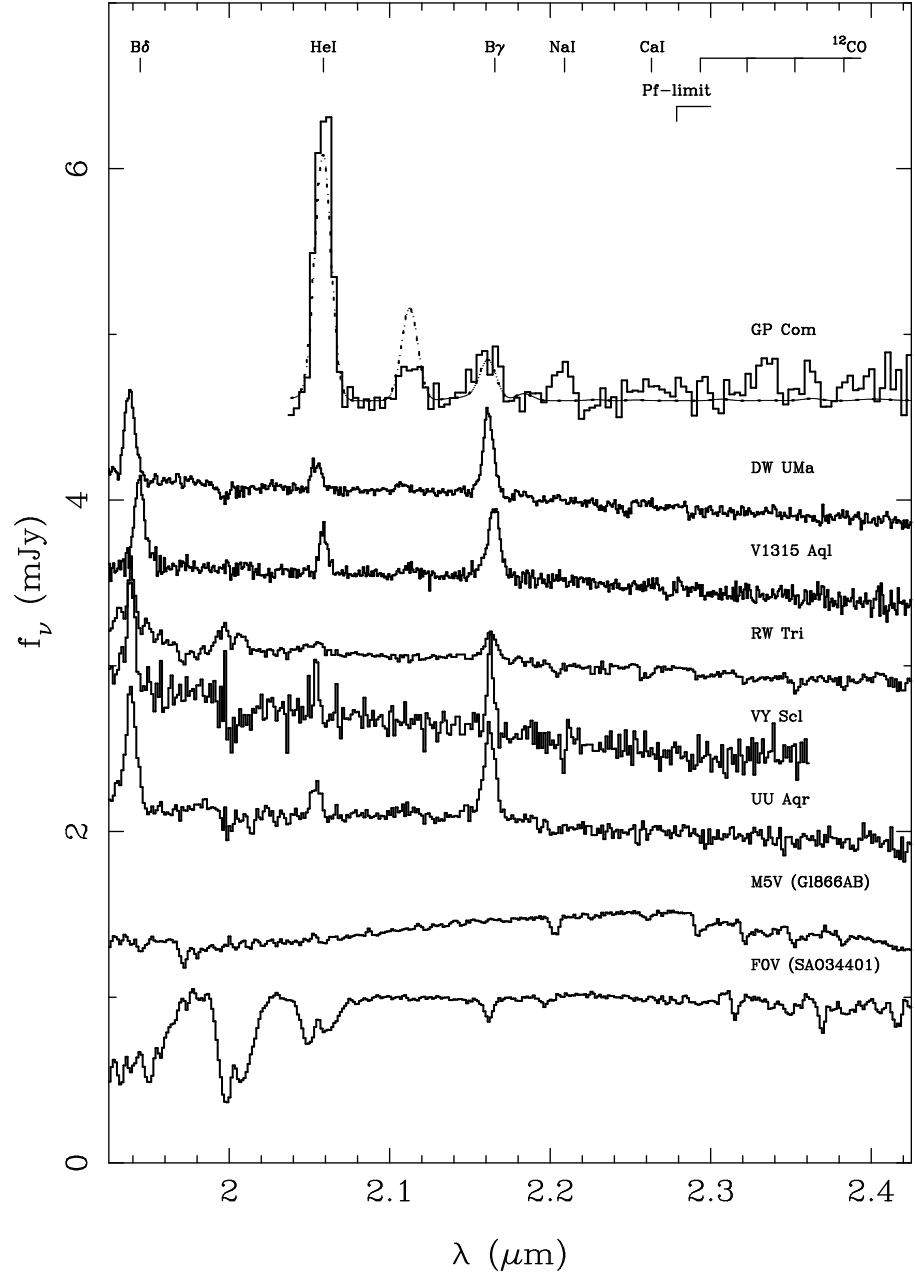}
%\vspace{6.6in}
\caption{Infrared spectra of the  novalike  variables GP Com, DW  UMa,
V1315 Aql, RW Tri,  VY Scl, UU  Aqr and an M5V  star. The spectra have
been normalized by dividing by the flux at 2.24~$\mu$m and then offset
by   adding a multiple  of 0.5  to each spectrum.   Also  shown is the
spectrum of an F0V star, normalized by dividing by a spline fit to its
continuum,  which  indicates    the location of    telluric absorption
features.   The dashed line  under the spectrum of GP  Com  is a model
spectrum from gas in LTE (see text for details).} 
\end{figure}

In Figure~3 we present IR spectra of the novalike variables GP Com, DW
UMa,  V1315    Aql, RW Tri,  VY Scl    and UU  Aqr   (Dhillon \& Marsh
1997). Note that the anti-dwarf novae  DW UMa and  VY Scl were both in
their high state when the spectra in Figure~3 were obtained.  With the
exception   of  the double-degenerate  system GP   Com,  which will be
discussed in more detail below, the spectra  in Figure~3 are dominated
by strong, single-peaked   emission lines of H\,{\small  I} (B$\gamma$
and  B$\delta$) and  He\,{\small I}  (2.06~$\mu$m).  This is  in stark
contrast to what one might expect from standard accretion disc theory,
which  predicts that emission  lines from high inclination discs, such
as  those in the  eclipsing systems DW UMa, V1315   Aql, RW Tri and UU
Aqr,    should appear double   peaked.  The  absence of  double-peaked
profiles in high-inclination novalikes is also observed in the optical
(e.g. Dhillon 1996) and is one of  the defining characteristics of the
so-called SW Sex stars, of which DW UMa and V1315 Aql are members. 

RW  Tri  is  the  only novalike   presented   in Figure~3 which  shows
absorption  features from the  secondary star --  one can clearly make
out the  profiles of Na\,{\small I}, Ca\,{\small   I} and $^{12}$CO in
the     spectrum.     The  distinctive    water   absorption   band at
$\sim$2.3~$\mu$m, so prominent in the spectrum of  the M5V star at the
foot of Figure~3,  is absent in  RW Tri, indicating that its secondary
is most likely a late K-dwarf.  This is confirmed  both by the orbital
period of RW Tri  (5.6~hr), for which one would  expect a secondary of
spectral type $\sim$K7--M0~$^2$ and by the skew mapping experiments of
Smith,  Cameron \&  Tucknott (1993).   There  is no  evidence for  the
secondary star in any of the other novalikes in Figure~3. This implies
that the  discs in these novalike  variables (which are all just above
the period gap) contribute a much larger fraction of the IR light than
the discs  of dwarf novae  just above  the period  gap  (e.g.  IP Peg,
Figure~1).  In  fact, it is possible  to place an  upper  limit on the
amount of  light   contributed  by  the   secondary star,   even  when
absorption features from the secondary  are not visible.  This can  be
done by normalizing all spectra and then  subtracting a constant times
a   spectral-type template from  the CV   spectrum and inspecting  the
residuals for any  (reversed) secondary star  features.  The value  of
the constant  at which these reversed features  become apparent in the
residual spectrum  then represents an  upper  limit to the  fractional
contribution of the secondary star  to the total light.  Applying this
technique to the  novalikes in  Figure~3,  we find that  the secondary
star generally contributes $\lta$25\%  to the K-band light.  A similar
upper limit to the secondary star contribution was deduced from I-band
observations of the novalike DW UMa in a low state by Marsh \& Dhillon
(1997).  This  is   remarkable, as the   low state  was some   3  to 4
magnitudes fainter than the normal state of  DW UMa, and yet there was
still no sign of the  secondary star. This  implies that the secondary
star  in  DW UMa has an   apparent magnitude of   I$>$19.5 and hence a
distance of at  least $\sim$850 pc  if the secondary star has spectral
type M4.$^2$  If this lower-limit to  the distance is typical  of most
novalikes (or specifically, SW Sex stars), then it means that the mass
transfer rates  derived   from techniques   such as eclipse    mapping
(Rutten, van Paradijs \&  Tinbergen 1992) are underestimating the true
values. 

GP Com consists of a CO white dwarf and a helium degenerate star in an
orbit  of 46~min  period. Neither  star   is directly  visible  -- the
optical  light  from the system  is dominated  by the  accretion disc,
which has  a spectrum composed almost entirely  of helium and nitrogen
emission lines  reflecting  the products of  hydrogen burning  and CNO
processing in the helium degenerate donor. A  very simple model, based
upon LTE   emission from an  $\sim 11\,000$~K  optically  thin (in the
continuum)  slab,  provides a  surprisingly   good fit to  the optical
spectrum of GP Com (Marsh, Horne \& Rosen 1991). In Figure~3, the same
model has been applied to the IR spectrum of GP Com, with equally good
results. The  model  predicts the existence  of three  strong emission
lines in the K-band, all  of He\,{\small I},  which are all present in
the  actual spectrum. Note  that  there  is also some  evidence  for a
fourth emission line in the  spectrum at 2.2~$\mu$m  that we have been
unable to identify. 

\subsection{Polars}

\begin{figure}
\psfig{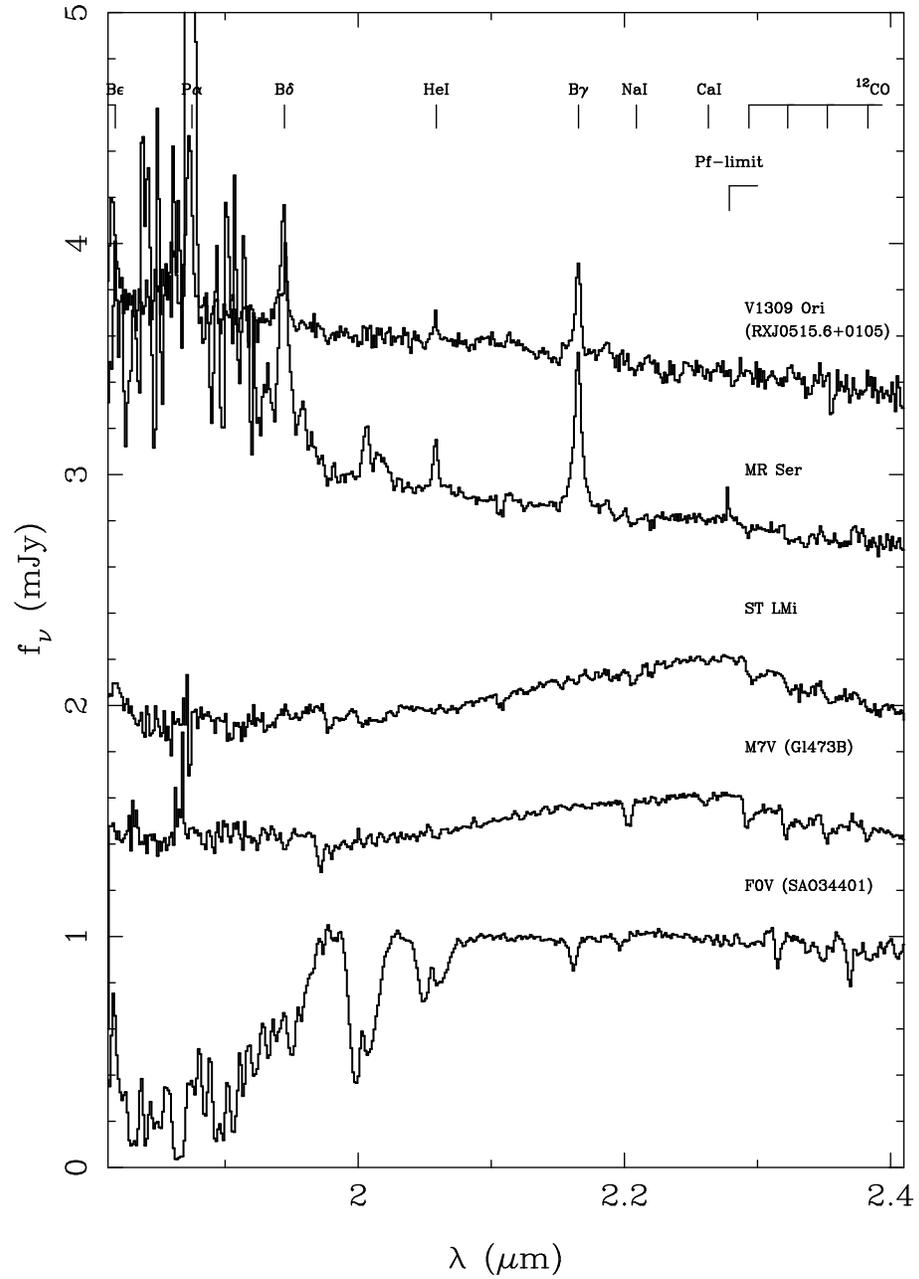}
%\vspace{6.8in}
\caption{Infrared spectra of the polars V1309 Ori,  MR Ser, ST LMi and
an M7 dwarf star. The spectra have been  normalized by dividing by the
flux at 2.24~$\mu$m  and then offset by  adding  a multiple of  0.6 to
each spectrum.  Also shown is the  spectrum of an F0V star, normalized
by dividing by   a spline fit to  its  continuum, which indicates  the
location of telluric absorption features.} 
\end{figure}

Polars were the first  class  of CV to   be studied in detail with  IR
spectroscopy (Bailey, Ferrario \& Wickramasinghe 1991). The motivation
behind  these  and  subsequent   observations   (Ferrario,  Bailey  \&
Wickramasinghe  1993,  1996) was  a desire  to  determine the magnetic
field strength  of  the white dwarf  by observing  cyclotron humps  in
their spectra.  Cyclotron radiation in polars is produced by electrons
gyrating along magnetic field lines in the accretion shock at the base
of  the accretion  column  and  occurs at  discrete harmonics  of  the
fundamental frequency. In the high-energy environment of the accretion
region, the harmonics are  broadened and merge,  causing humps in  the
emitted   spectrum. The  shape    of  these cyclotron    humps provide
invaluable  measurements   of the   magnetic field  strength, density,
temperature and optical depth in the emission region. 

With fields in  the range $B\sim$10--70~MG, the  fundamental cyclotron
frequencies in polars are in the IR from approximately 10 to 1 $\mu$m,
respectively.     The   cyclotron humps   extend   bluewards  from the
fundamental,  the higher  order harmonics  generally  appearing in the
optical  and the lower order  harmonics in the  IR.  It is not easy to
detect high-order harmonics as they are  often smeared by field spread
and variations in temperature and density  across an extended emission
region.   Low-order harmonics, however,   are more  easily resolvable,
which led Bailey, Ferrario and Wickramasinghe  to obtain IR spectra of
the polars AM Her,  ST LMi, EF Eri  and BL Hyi and successfully detect
cyclotron humps in these  relatively weak-field systems for  the first
time. 

In  Figure~4 we  present   IR spectra of  V1309  Ori  (RXJ0515.6+0105;
Harrop-Allin et al. 1997a), MR Ser and ST  LMi (Dhillon et al. 1997b).
V1309 Ori  has a 7.98-h orbital period,  the longest of any  polar yet
identified. The spectrum in Figure~4 was obtained around orbital phase
0.5 and reveals strong  Paschen, Brackett and He\,{\small  I} emission
lines from the accretion column. No obvious spectral features from the
secondary star are detected; the K-band  emission is dominated instead
by cyclotron  radiation  from the accretion  region.  By modelling the
cyclotron emission in V1309 Ori, Harrop-Allin et al. (1997a) concluded
that  the  shape    of the  IR   spectrum   is   consistent   with the
long-wavelength tail of the cyclotron fundamental from a 60 MG field. 

The  IR spectrum  of  MR Ser shows no  evidence  of cyclotron humps or
secondary star features. ST LMi, on the other hand, appears to have an
IR spectrum composed entirely of light from the  secondary star, as it
looks  virtually  identical to  the spectrum of   the M7V star plotted
directly beneath  it.  This is close   to the spectral type  one might
expect of  the secondary star given  the 1.90-hr orbital  period of ST
LMi.$^2$ One can see that the rapid rotation of  the secondary star in
ST LMi has  broadened the $^{12}$CO absorption  bands when compared to
the $^{12}$CO profiles in the slowly rotating M7 field dwarf. The most
likely explanation for the spectrum of ST LMi is that we have observed
it during one of its low states when  there is no accretion occurring.
It is exceptional, however, for the emission lines from the irradiated
inner hemisphere of the secondary star to disappear in polars during a
low state -- the obvious explanation that our spectrum was obtained at
an orbital phase when the inner hemisphere was obscured cannot be true
as the spectrum in Figure~4 is a whole-orbit average. 

\subsection{Intermediate polars}

\begin{figure}
\psfig{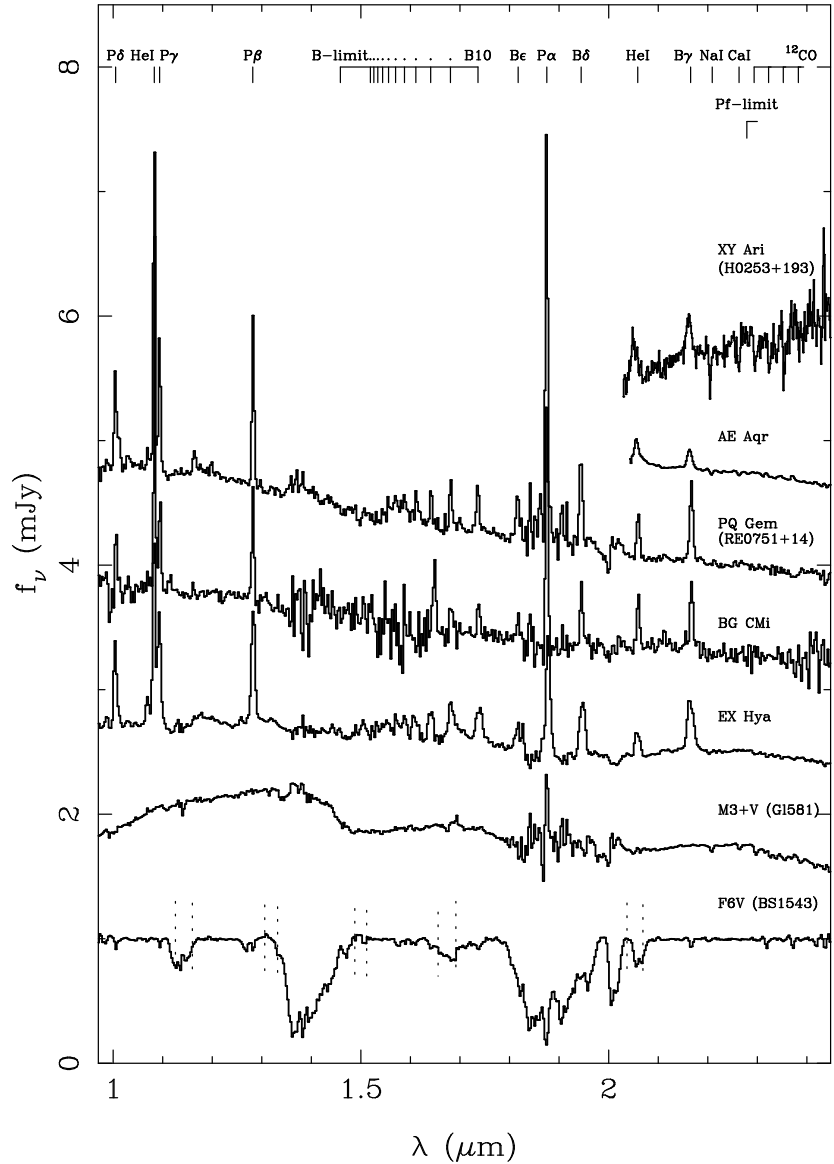}
%\vspace{6.0in}
\caption{Infrared spectra  of the intermediate  polars XY Ari, AE Aqr,
PQ Gem, BG CMi,  EX Hya and an  M3+ dwarf star.  The spectra have been
normalized by dividing  by the flux at 2.24~$\mu$m  and then offset by
adding a multiple  of 0.75 to each  spectrum (with the exception of XY
Ari, which has been normalized by dividing by 2 (effectively expanding
its  $y$-axis by a  factor of 3.7  relative  to the other spectra) and
then offset by adding 2.1). Also shown is the spectrum of an F6V star,
normalized by  dividing  by  a spline  fit   to  its continuum,  which
indicates the location of telluric absorption features. The spectra of
PQ Gem, BG CMi and EX Hya  consists of six  subspectra which have been
merged by matching the flux levels in overlapping regions (illustrated
by the vertical dotted lines in the lowermost spectrum).} 
\end{figure}

The intermediate  polars (IPs) are   believed to have  magnetic  field
strengths   intermediate between  polars and dwarf   novae.  The white
dwarf rotation  is  not synchronized with  the  orbit in IPs,  and the
rotating magnetosphere of the primary is able to disrupt the accretion
disc  out to a radius   where the ram pressure  of  the accretion flow
balances the magnetic pressure: sufficiently strong fields are able to
prevent the formation of an accretion disc  altogether. A knowledge of
the magnetic  field  strength  in  IPs  is thus   vitally important in
determining the mode of accretion in IPs. Unfortunately, it has proved
remarkably difficult to   determine the field strength  in  IPs due to
their weak fields.  The success of Bailey, Ferrario and Wickramasinghe
in measuring cyclotron features in the IR spectra of weak-field polars
prompted Dhillon et al.  (1997a)  to observe intermediate polars (IPs)
in the IR, since if the  IPs have weaker  fields than polars one would
expect their cyclotron  humps to  be  most easily observed  in the IR,
where the  cyclotron humps are closer  to  their fundamental frequency
and therefore more easily resolvable.  The  results of this study  are
presented in Figure~5,  where we show  IR spectra  of the intermediate
polars XY Ari  (Harrop-Allin et al.  1997b), AE Aqr (Dhillon  \& Marsh
1995), PQ Gem, BG CMi and EX Hya (Dhillon et al. 1997a). 

The  spectra of  the  IPs in  Figure~5  are all  dominated  by strong,
single-peaked emission lines of H\,{\small  I} and He\,{\small I} from
the    accretion regions.  Absorption    features of   Na\,{\small I},
Ca\,{\small I}  and  $^{12}$CO from   the secondary  star are  clearly
observed in XY Ari  and AE Aqr.  The K4V  secondary star in AE Aqr has
already  been well studied  in the optical,  but the IR spectrum of XY
Ari  and  the  detection of  its secondary  are   unique, owing to its
location behind a dense molecular  cloud which renders it invisible in
the optical (Zuckerman et al. 1992).  The  strong reddening is evident
in the continuum of  XY Ari, as is  the  absence of a change  in slope
around 2.3~$\mu$m, implying that the secondary is most likely a K-star
(in agreement with the orbital period of 6.06~hr~$^2$).  The secondary
star has also been detected in the IR spectrum of EX Hya, in which the
continuum  is dominated by water  absorption bands around 1.4, 1.7 and
2.3~$\mu$m  from an $\sim$M3V secondary.   In none  of the spectra  in
Figure~5 is  there any evidence  for cyclotron  humps and hence  it is
impossible to measure the magnetic   field strengths of the IPs   from
these data  (Dhillon et al.  1997a).   It  may  be possible to  detect
cyclotron humps in IPs with higher  signal-to-noise IR spectra as long
as  great  care is  taken correcting  for  telluric absorption and the
secondary star spectrum (which is also humpy in appearance).  A better
approach  would be to use IR  circular spectropolarimetry, which would
be sensitive only  to cyclotron emission  and hence insensitive to the
contributions of the secondary star and telluric absorption.

\end{document}